\documentclass{aa}
\usepackage[dvips]{graphicx}
\usepackage{natbib}
\usepackage{txfonts}

\newcommand{\hii}{\hbox{H\,\textsc{ii}}}
\newcommand{\ha}{\ensuremath{{\textrm{H}\alpha}}}
\newcommand{\hb}{\ensuremath{{\textrm{H}\beta}}}
\newcommand{\oii}{[\textrm{O}\,\textsc{ii}]}
\newcommand{\oiii}{[\textrm{O}\,\textsc{iii}]}

\newcommand{\lam}{\ensuremath{\lambda}}

\newcommand{\unit}[2][]{\ensuremath{\,\textrm{#2}^{#1}\,}}
\newcommand{\x}{$\times$}
\newcommand{\ab}{$\sim$}
\def\asec{$''$\hspace*{-.1cm}}
\def\lam{$\lambda$}
\newcommand{\lsun}{\ensuremath{L_\odot}}

\begin{document}

\title{Low-excitation blobs in the Magellanic Clouds\thanks{Based 
on observations obtained at the European Southern Observatory, 
La Silla, Chile} 
}

\offprints{Fr\'ed\'eric Meynadier, \hspace{1cm} \\Frederic.Meynadier@obspm.fr}

\date{Received .../Accepted ...}

\titlerunning{Low-excitation blobs}
\authorrunning{Meynadier \& Heydari-Malayeri}

\author{F. Meynadier\inst{1,2} \and M. Heydari-Malayeri\inst{1}}

\institute{LERMA, Observatoire de Paris, 61 Avenue de l'Observatoire, 
F-75014 Paris, France \and UPMC -- Universit\'e Paris 6, 4 place Jussieu, 75005 Paris, France}

\abstract
{}
{We study an unknown, or very poorly known, interstellar \hii\, component in 
the Magellanic Clouds. This is the first study ever devoted to this class of objects, which 
we call Low-excitation blobs (LEBs). 
 }
{We used low-dispersion spectroscopy carried out at ESO to obtain
  emission line intensities of \ha, \hb, and \oiii\ (\lam \lam\,4959\,+\,5007) 
for 15 objects in the Large Magellanic Cloud and 
14 objects in the Small Magellanic Cloud. Results are
  displayed in  excitation (\oiii/\hb\ ratio) versus \hb\ luminosity
  diagrams.
}
{ We show the presence of an LEB component in the Magellanic Clouds and study its  
relationship with the already known class of high-excitation blobs (HEBs). 
The newly found LEBs are lower 
excitation counterparts of 
HEBs and are powered by less massive exciting stars. Further study of LEBs 
is expected to provide new pieces of information for a better understanding the low mass end of 
the upper initial mass function in the Magellanic Clouds.  
}
{}


\maketitle


\section{Introduction}

The ionized content of the Magellanic Clouds (MCs) is not limited 
to giant and supergiant \hii\, regions or to a relatively small 
number of supernova 
remnants. Detection of high-excitation \hii\, blobs (hereafter HEBs)
showed the presence of an unknown \hii\, component lying mainly  
 adjacent to or toward the typical giant \hii\, regions, and they are rarely
in isolation 
\citep[see] [ and references therein]{mhm02a}.  
In contrast to the ordinary \hii\, regions
of the MCs, which are extended structures spanning
several arcminutes on the sky ($>$ 50 pc) and are powered by a
large number of hot stars, HEBs are very dense small regions usually
4\asec\, to 10\asec\, in diameter (1 to 3 pc) affected by local dust. 
Here we aim at highlighting  another almost unknown or very poorly 
known constituent of the interstellar medium in the MCs, which we call 
low-excitation blobs (LEBs). So far, no detailed analysis has been 
devoted to these objects, which outnumber the high-excitation ones 
although  \cite{sand-phil}, in their search for 
planetary nebulae, presented a list of three low-excitation compact 
nebulae in the SMC. Similarly, \cite{morgan} detected six 
additional objects of this type, which have \oii\,\lam 3727 lines stronger 
than the \hb\, and the doublet of \oiii\,\lam\lam 4959, 5007 weak compared to 
\hb. \\

The study of LEBs is interesting for several reasons. 
It sheds light on a genuine ionized component of the interstellar medium 
in the MCs. Moreover, since LEBs are excited by massive stars of 
lower temperature/mass, compared with those that power HEBs and 
typical giant 
\hii\, regions, investigating them will provide useful information 
for better understanding the low-mass end of the upper initial mass 
function. 
In particular, they hint at the physical conditions necessary for 
massive star formation in the relatively small molecular clouds that are 
their likely birth places.  
Therefore, LEBs represent a thus far ignored interface between several 
aspects of massive star formation and the interstellar environment.

\section{Observations}

The observations were performed from 12 to 17 October 1989, with the
ESO 1.52m telescope at La Silla (Chile). 
A Boller \& Chivens spectrograph was used with grating \#23, which had a dispersion 
of 114\,\AA\,mm$^{-1}$ and ranged in wavelength  
from 4772 to 6710\,\AA , centered on 5741\,\AA. 
The detector, CCD \#13 (type RCA SID 006 EX), had 
1024\,\x\,640 pixels of size 15\,\x\,15 $\mu$m, each pixel corresponding 
to 0\asec.68 on the sky; it was binned 2\,\x\,2 pixels. 
The resulting data were frames of 515\,$\times$\,101 pixels covering
$\sim$\,69\asec\, on the sky along the slit direction. 
The spatial scales per arcsec for LMC and SMC are 0.24 and 0.32 pc, 
respectively.
\\

\begin{figure}
\includegraphics[width = \linewidth]{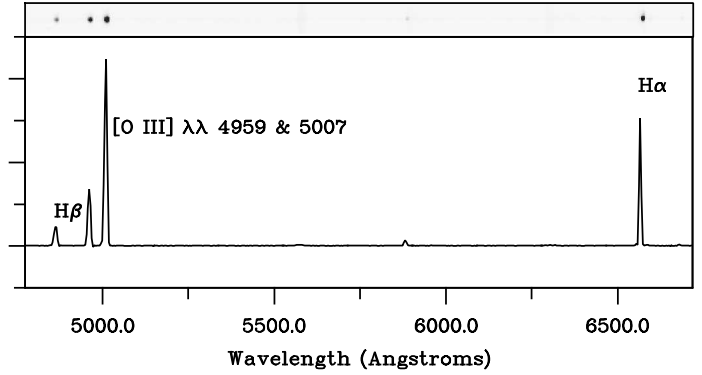}
\caption{A sample spectrogram belonging to SMC N\,88A. The upper panel
  shows the 2D spectrum and the lower panel a cut through the central
  row, with relative intensity scale. 
}
\label{fig:spect}  
\end{figure}

For all objects, we chose an entrance slit of 10\asec\, width in order
to secure the whole flux from the object. This was at the expense of
degrading the spectral resolution, which was not critical for our
purposes.  The nights were photometric, and several standard stars
were observed each night for flux calibration.  Care was taken to center the
objects in the middle of the 10\asec-width slit, in order to ensure
the measurement of the whole flux. An example of the spectra is shown in 
Fig. \ref{fig:spect}. \\

Targets were selected from the \citet{henize56} catalog on the 
basis of two criteria: 1) \ha\, emission, and 
2) small angular size (a few arcseconds). 
Subsequently,  
a number of them turned out to be planetary nebulae (Table
\ref{tab:resultats_misc}),
as explained in Sect. \ref{sec:discussion}. It should be underlined that the sample is not
complete, mainly as far as the LEBs are concerned. Several other fainter 
LEBs are expected to be present in other MC nebular surveys, for example, 
\cite{davies76}.
  A summary of the observations is listed in
Table \ref{tab:obs_log}.  Some of the objects were observed
several times in order to enhance the S/N ratio.

\begin{table*}
\begin{center}
\caption{Log of observations (exposure times indicated in parenthesis)
\label{tab:obs_log}
}
\begin{tabular}{lp{.4\linewidth}p{.4\linewidth}}

Date & LMC objects & SMC objects\\
\hline\hline
12/10 & 
N\,11A (2 min), 
N\,160A1 (2 min) &
N\,9 (5 min),
N\,11 (2 min),
N\,77 (5 min),
N\,81 (5 min)\\
\hline
13/10 &
N\,11A (2 min), 
N\,83B (4$\times$1 min),
N\,156 (8 min),
N\,177 (6 min),
N\,159 (4 min) & 
N\,9 (5 min),
N\,45 (5 min),
N\,55 (5 min),
N\,63 (5 min),
N\,64 (5 min),
N\,77 (8 min),
N\,81 (1 min),
N\,88A (40 s) \\
\hline
14/10 &
N\,68 (5 min),
N\,90 (10 min),
N\,191A (2 min),
N\,193 (3 min) &
N\,1 (5 min),
N\,61 (5 min),
N\,66 (5 min),
N\,70 (10 min) \\
\hline
15/10 &
N\,66 (5 min),
N\,103 (10 min), 
N\,105A-IR (11 min)&
N\,10 (10 min),
N\,11(5 min),
N\,21 (5 min),
N\,26 (2 min),
N\,73 (10 min),
N\,75 (10 min),
N\,78 (10 min)\\
\hline
16/10 &
N\,33 (5 min),
N\,44 (2 min),
N\,66 (5 min),
N\,68 (5 min),
N\,82 (1 min),
N\,88 (1 min),
N\,197 (10 min) &
N\,29 (10 min),
N\,31 (10 min),
N\,32 (10 min),
N\,33 (3 min),
N\,47 (4 min),
N\,63 (5 min),
N\,64 (5 min) \\
\hline
17/10 &
N\,6 (20 min),
N\,11E (5 min),
N\,159-5 (3 min)&
N\,68 (5 min),
N\,81 (1 min),
N\,88A (1 min)\\
\hline
\end{tabular}
\end{center}
\end{table*}

\subsection{Data reduction}

The data processing from raw images to calibrated 2D spectra was done
using standard \texttt{iraf} procedures. Flux calibration was
performed with 7, 7, 5, 6, 5, and 4 spectrophotometric
standard stars respectively for the nights between 12 and 17 October. \\

A semi-automated \texttt{pyraf} task was written in order to
retrieve the flux from the 2D data. It allowed us to integrate the
flux in each line while defining what zones were usable for background
estimation and removing point-like source fluxes. In some cases, that 
objects did not fit into the 10\asec\ \ slit may lead to a small 
under-estimation of the flux in every line.

\subsection{Error estimate}

\label{sec:error}

The errors involved in the absolute fluxes were estimated 
assuming a Poisson law for the signal.  The S/N, 
expressed as  $F/\Delta F$ where $F$ is the line flux  and
$\Delta F$ the corresponding uncertainty, should be proportional to
$\sqrt{N}$, where $N$ represents the number of collected photons. With $N$
itself expressed as $F \times t$ (where $t$ is exposure time), the 
S/N can therefore be written as:

\begin{equation}
\frac{F}{\Delta F} = k \sqrt{Ft}.
\label{eq:1}
\end{equation}

\hspace*{-.7cm}
The coefficient $k$ is a constant, possibly different for each line,
which we determine from the SMC N\,81 measurements. This object was 
observed on three different nights, and for each line (\ha,
\oiii$\lambda$4959, \oiii$\lambda$5007, \hb) we calculate:

\begin{equation}
k_\lambda = \frac{\sqrt{F_\lambda}}{\Delta F_\lambda \sqrt{t} }
\end{equation}

\hspace*{-.7cm}
where $F_\lambda$ is the average flux over the three nights, $\Delta
F_\lambda$ the spread of the measurements (around 20\% of the average flux),  
and $t$ the mean exposure time. 
Then, $k_\lambda$ is used in Eq. \ref{eq:1} to estimate the relative
uncertainty on absolute flux measurements for other objects. \\

\begin{figure*}
\includegraphics[width = .84\linewidth]{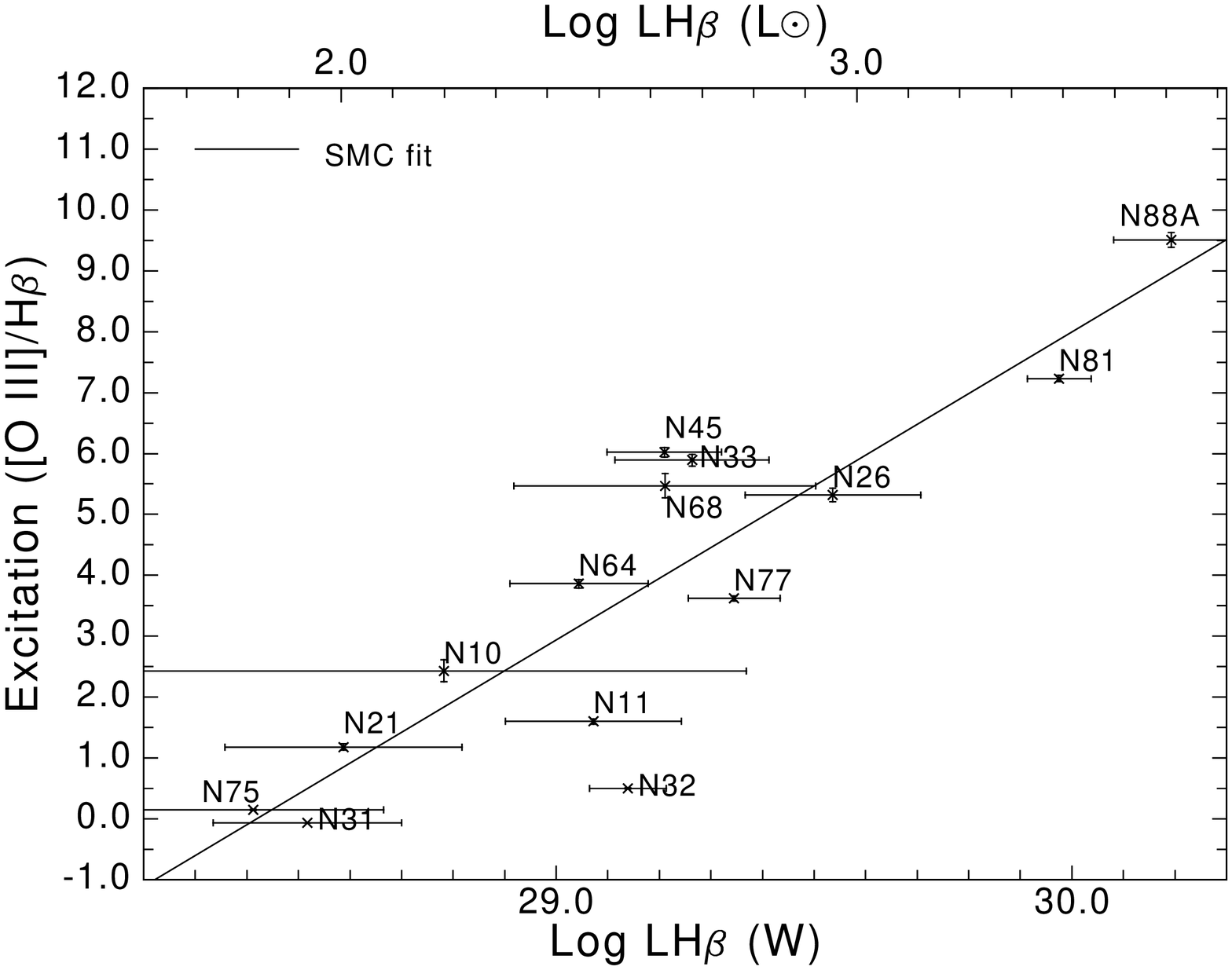}
\includegraphics[width = .84\linewidth]{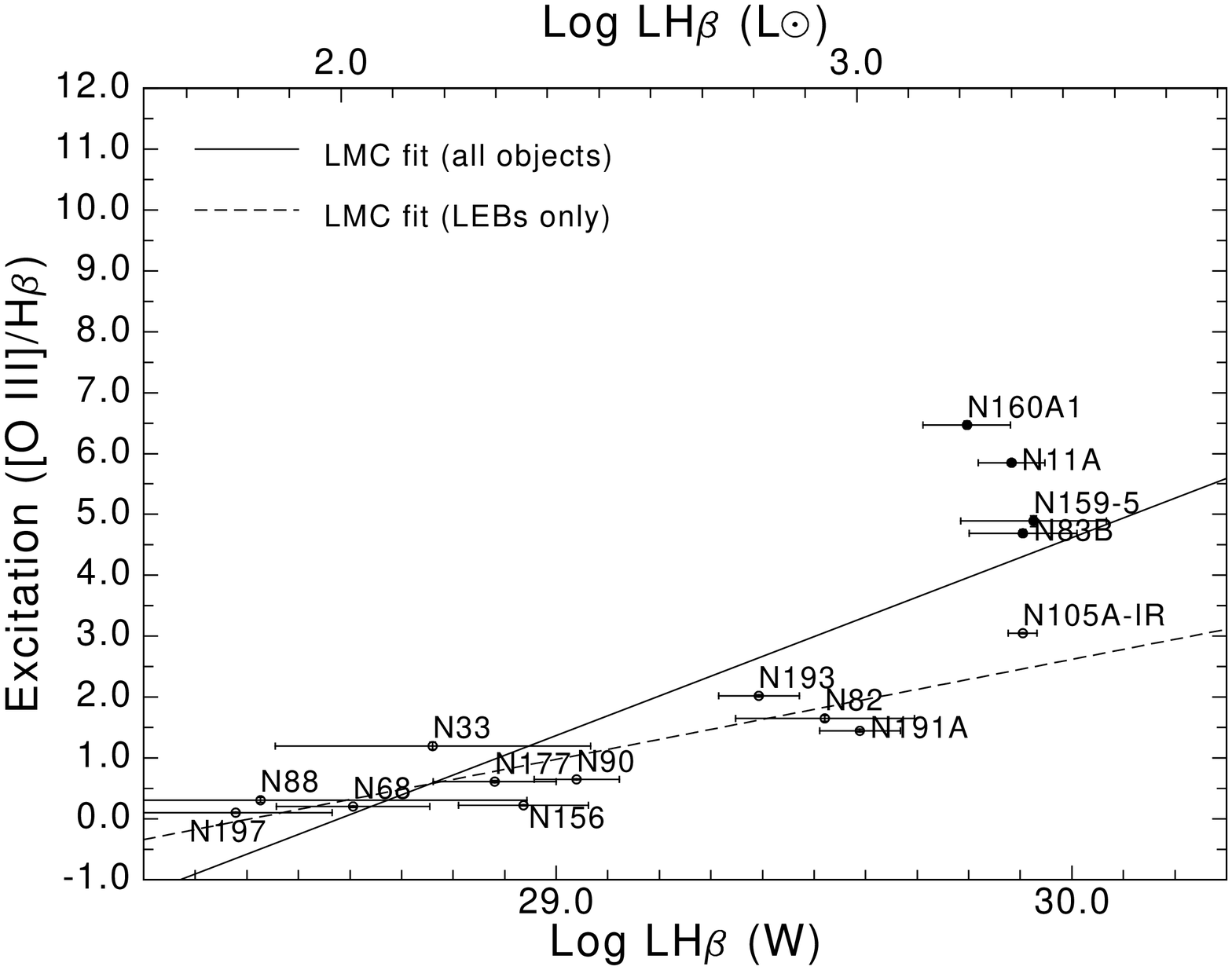}
\caption{
Excitation (measured by the line ratio \oiii\ \lam \lam4959\,+\,5007/\hb)
versus the \hb\ luminosity for the SMC and LMC objects. 
The logarithmic luminosity scale is in units of watts W, $10^7$ erg s$^{-1}$ 
as well as in units of solar bolometric luminosity. 
The figures are based on the data presented in Tables
\ref{tab:resultats_SMC} and \ref{tab:resultats_LMC}, respectively. 
Solid lines: least-square fits for the SMC and LMC objects.
Dashed line: least-square fit for the LMC sample, excluding HEBs (filled dots). 
}
\label{fig:diag}  
\end{figure*}

Line ratios are more precisely known, because  
they are (to the first order) insensitive to calibration errors and
sky transparency
variations from night to night; \oiii/\hb\ and \ha/\hb\ ratios 
show only a $\sim$ 1\% variation. To take this fact into account, the
error estimate for line ratios was derived from a set of data
taken during the same night on LMC N\,83B. The data's spread is
representative of the non-systematic errors, and we use it to
calculate new $k_\lambda$ coefficients that are adapted to relative
photometry error estimates. 
The uncertainty on the $\log(L\hb)$ and \oiii/\hb\ values are
graphically displayed in Fig. \ref{fig:diag} as error bars.

\section{Results}

\begin{table*}[f]
\begin{footnotesize}
\begin{center}
\caption{LMC line measurement results
\label{tab:resultats_LMC}
}
\begin {tabular}{lrrrrrrrrr}
\hline
\hline
Object & F(\hb) & \oiii\, 4959+5007&\ha  & c(\hb) & \oiii / \hb  & I(\hb) &
L(\hb) & radius & $<$Ne$>$\\ 
LMC & 
$10^{-16}$&
$10^{-16}$&
$10^{-16}$&
&
&
$10^{-16}$&
$10^{26}$&
pc&
\unit[-3]{cm}
\\
&\unit{W}/\unit[2]{m}  &
\unit{W}/\unit[2]{m}  &
\unit{W}/\unit[2]{m}  &
&
&
\unit{W}/\unit[2]{m}  &
\unit{W} &&
\\
\hline
         N\,11A &   85.8 &  518.0 &  339.7 &   0.46 &   5.85 &  253.0         &   7638 &  0.7 &   1170\\
          N\,33 &    1.6 &    2.1 &    9.9 &   1.07 &   1.20 &   19.1         &    576 &  0.2 &   1650\\
          N\,68 &    6.7 &    1.4 &   23.7 &   0.30 &   0.20 &   13.4         &    404 &  1.3 &    110\\
          N\,82 &   24.5 &   42.3 &  110.6 &   0.65 &   1.65 &  109.9         &   3319 &  0.2 &   4260\\
         N\,83B &   66.6 &  326.8 &  290.5 &   0.60 &   4.69 &  266.4         &   8043 &  0.3 &   3710\\
          N\,88 &    3.4 &    1.1 &   13.0 &   0.42 &   0.31 &    8.8         &    267 &  1.0 &    130\\
          N\,90 &   10.6 &    7.2 &   44.0 &   0.54 &   0.65 &   36.3         &   1096 &  1.4 &    160\\
     N\,105A-IR &   84.9 &  268.6 &  344.0 &   0.50 &   3.05 &  265.9         &   8028 &  1.1 &    690\\
         N\,156 &    5.7 &    1.3 &   26.7 &   0.70 &   0.22 &   28.6         &    863 &  1.0 &    240\\
       N\,159-5 &   11.8 &   63.9 &   88.5 &   1.37 &   4.89 &  279.3         &   8431 &  0.6 &   1600\\
       N\,160A1 &   50.7 &  342.9 &  222.6 &   0.61 &   6.47 &  207.1         &   6253 &  0.6 &   1530\\
         N\,177 &    8.6 &    5.4 &   34.1 &   0.47 &   0.61 &   25.2         &    760 &  0.8 &    320\\
        N\,191A &   60.2 &   89.2 &  216.9 &   0.33 &   1.45 &  128.6         &   3883 &  0.7 &    960\\
         N\,193 &   39.9 &   82.5 &  142.0 &   0.31 &   2.02 &   81.9         &   2472 &  0.9 &    500\\
         N\,197 &    2.1 &    0.2 &    9.1 &   0.57 &   0.10 &    7.9         &    239 &  1.0 &    130\\

\hline
\end{tabular}
\end{center}
\end{footnotesize}
\end{table*}

\begin{table*}[f]
\begin{footnotesize}
\begin{center}
\caption{SMC line measurement results
\label{tab:resultats_SMC}
}
\begin {tabular}{lrrrrrrrrr}
\hline
\hline
Object & F(\hb) & \oiii\, 4959+5007&\ha  & c(\hb) & \oiii / \hb  & I(\hb) &
L(\hb)& radius & $<$Ne$>$\\ 
SMC & 
$10^{-16}$&
$10^{-16}$&
$10^{-16}$&
&
&
$10^{-16}$&
$10^{26}$&
pc&
\unit[-3]{cm}
\\
&\unit{W}/\unit[2]{m}  &
\unit{W}/\unit[2]{m}  &
\unit{W}/\unit[2]{m}  &
&
&
\unit{W}/\unit[2]{m}  &
\unit{W} &&\\
\hline
          N\,10 &    2.7 &    7.0 &   12.3 &   0.64 &   2.43 &   11.9         &    605 &  3.0 &     40\\
          N\,11 &    9.1 &   14.7 &   34.5 &   0.40 &   1.60 &   23.2         &   1180 &  1.0 &    270\\
          N\,21 &    2.9 &    3.5 &   11.1 &   0.42 &   1.18 &    7.6         &    387 &  0.6 &    360\\
          N\,26 &   12.7 &   71.2 &   60.4 &   0.73 &   5.32 &   67.8         &   3440 &  0.5 &   1350\\
          N\,31 &    2.2 &   -0.1 &    8.8 &   0.46 &  -0.06 &    6.5         &    329 &  0.4 &    660\\
          N\,32 &   13.1 &    6.7 &   46.8 &   0.32 &   0.50 &   27.1         &   1377 &  1.4 &    190\\
          N\,33 &   11.0 &   67.6 &   45.3 &   0.51 &   5.89 &   36.1         &   1834 &  0.5 &    920\\
          N\,45 &   11.9 &   73.9 &   45.9 &   0.43 &   6.02 &   31.9         &   1621 &  1.4 &    190\\
          N\,64 &    8.2 &   32.8 &   31.7 &   0.42 &   3.86 &   21.8         &   1107 &  1.5 &    150\\
          N\,68 &    1.8 &   10.6 &   12.2 &   1.26 &   5.47 &   32.0         &   1625 &  0.4 &   1380\\
          N\,75 &    1.2 &    0.2 &    5.3 &   0.63 &   0.15 &    5.1         &    259 &  2.6 &     30\\
          N\,77 &   14.9 &   55.7 &   59.1 &   0.47 &   3.62 &   43.6         &   2214 &  1.7 &    180\\
          N\,81 &  124.9 &  915.4 &  403.3 &   0.17 &   7.23 &  186.1         &   9451 &  1.1 &    750\\
         N\,88A &   88.0 &  871.7 &  368.5 &   0.54 &   9.51 &  307.1         &  15594 &  0.5 &   2620\\

\hline
\end{tabular}
\end{center}
\end{footnotesize}
\end{table*}

\begin{table*}[f]
\begin{footnotesize}
\begin{center}
\caption{Objects excluded from the HEB/LEB sample 
}
\label{tab:resultats_misc}
\begin {tabular}{lrrrrrrrrl}
\hline
\hline
Object & F(\hb) & \oiii\, 4959+5007&\ha  & c(\hb) & \oiii / \hb  & I(\hb) &
L(\hb) & radius & Notes\\ 
 & 
$10^{-16}$&
$10^{-16}$&
$10^{-16}$&
$10^{-16}$&
&
$10^{-16}$&
$10^{26}$& 
pc&
\\
&\unit{W}/\unit[2]{m}  &
\unit{W}/\unit[2]{m}  &
\unit{W}/\unit[2]{m}  &
\unit{W}/\unit[2]{m}  &
&
\unit{W}/\unit[2]{m}  &
\unit{W} &
&
\\
\hline
&&&&&&&\\
\multicolumn{8}{l}{Planetary Nebula, LMC :}&\\
N\,66 &   3.5 &  40.2 &  11.4 &  11.37 &  0.17 &   5.2 &    157 & 0.3 &(a) (b) \\
\hline\\
\multicolumn{8}{l}{Symbiotic star, SMC :}\\
N\,73 &   0.7 &   0.1 &   7.2 &   0.20 &  1.89 &  51.5 &   2615 &0.3 &(c) \\
\hline\\
\multicolumn{8}{l}{Planetary Nebulae, SMC :}\\
N\,1 &   2.5 &   9.1 &   9.8 &   3.64 &  0.45 &   7.1 &    361 &0.3 &(d) \\
N\,9 &  18.0 &  69.3 &  51.6 &   3.84 &  0.00 &  18.1 &    917 &0.7&(e) \\
N\,29 &   0.3 &   1.8 &   1.3 &   6.29 &  0.60 &   1.2 &     60 &0.3& (d) \\
N\,47 &   3.9 &  16.3 &  12.4 &   4.18 &  0.15 &   5.5 &    280 & 0.3&(d) \\
N\,61 &  10.0 &  22.4 &  35.7 &   2.25 &  0.31 &  20.6 &   1047 & 0.5& (f) \\
N\,70 &   2.9 &  18.1 &  11.7 &   6.24 &  0.50 &   9.1 &    462 & 0.2&(d) \\

\hline
\end{tabular}
\end{center}
(a)~\citep{pena04}.
(b)~\citet{morgan84}
(c)~\citep{walker83}
(d)~\citet{sanduleak78}
(e)~\citet{meyssonnier93}
(f)~\citet{jacoby02}
\end{footnotesize}
\end{table*}

The results of flux measurements in various spectral lines are
presented in Table \ref{tab:resultats_LMC} for the LMC blobs and Table
\ref{tab:resultats_SMC} for the blobs situated in the SMC. Similarly,
Table \ref{tab:resultats_misc} summarizes the results for those objects
in both clouds that turned out to be planetary nebulae rather than
blobs.  In those tables, column 1 lists the object identification
according to \citet{henize56}, column 2 presents the measured
uncorrected \hb\, flux, and column 3 the sum of 
uncorrected 
\oiii\, line fluxes
\lam\lam4959\,+\,5007.  Column 4 gives the \ha\, flux, which includes
the [N\textsc{ii}] doublet \lam \lam 6548, 6584. 
  The logarithmic  reddening coefficient at \hb, based on the 
  Balmer decrement, is presented in column 5; it was derived from  
  comparing the observed  \ha/\hb\, ratio with the theoretical ratio computed 
  by \cite{brocklehurst71} for $T_{e}= 10^{4}$ K and  $N_{e}= 10^{2}$ cm$^{-3}$ 
  under case B conditions. Column 6 lists the de-reddened line ratio 
  \oiii\, (\lam\lam4959\,+\,5007)/\hb\, and column 7 the de-reddened \hb\, flux. 
  Column 8  
gives the \hb\, luminosity based on distance moduli of
18.5 mag \citep{alves04} and 19.07 mag \citep{abrahamyan04} 
for LMC and SMC, 
  corresponding to \ab\,50 and \ab\,65 kpc,
respectively. Column 9 lists the 
diameter of the object in pc, derived from \ha.
More specifically, the \ha\, line profile was integrated into a one-pixel column 
in the dispersion direction; the result  was then fitted by a 
Gaussian, the FWHM of which yielded the radius. 
Finally, column 10 presents the average electron densities 
   derived from the formula

   \begin{equation}
   <N_{e}> = 5.10\times\,10\,^{4}\,D\,R\,^{-1.5}\,T_{e}\,^{0.44}\,I(\hb),
   \end{equation}

\noindent
which applies to a Str\"omgren sphere of pure hydrogen. There, $D$ is the distance 
   in kpc, $R$ the radius in pc, $T_{e}$ the electron temperature in K (assumed to 
   be 10$^{4}$ K), and {\it I(\hb )} the nebular flux in units of erg cm$^{-2}$ s$^{-1}$.\\

Figure \ref{fig:diag} displays the corresponding positions in an
excitation/luminosity diagram. The excitation is quantified by the
\oiii/\hb\ ratio, and the luminosity is derived from the de-reddened
\hb\ flux. We see that the distribution of SMC points differs from
that of LMC data. More specifically, blobs of comparable luminosity
tend to be slightly more excited if situated in the SMC. Moreover,
several LMC blobs are vertically piled up around a luminosity of
$10^{30}$ W, with \oiii/\hb\, ratios ranging from 3.0 to 7.0 (LMC
N\,160A1, N\,11A, N\,159, N\,83B, N\,105A). \\

The SMC and LMC samples have been fitted by least-square lines, as follows:

\begin{itemize}
\item SMC blobs ($\chi^2 = 24.9$): 
\begin{equation}
\frac{\oiii}{\hb} = 5.1 (\pm 0.8) \log{(L_\hb)} - 144(\pm 23)
\label{eq:fitsmc}
\end{equation}
\item LMC blobs ($\chi^2 =  20.0$):
\begin{equation}
\frac{\oiii}{\hb} = 3.2 (\pm 0.6) \log{(L_\hb)} - 93(\pm 17).
\label{eq:fitlmc}
\end{equation}
\end{itemize} 

The goodness of the SMC fit is affected by N\,32: without
this point, the $\chi^2$ coefficient falls to 16.6 but the coefficients remain
unchanged. On the other hand, fitting the LMC data by a single
line is not satisfactory: the vertical accumulation of several blobs
around $L(\hb)=10^{30}$ W would suggest dropping them from the
least-square fit (see Sect. 4). This leads to the following
result:

\begin{itemize}
\item Least-square fit for LMC objects, excluding N\,160A1, N\,11A,
  N\,159-5, and N\,83B ($\chi^2 =  1.9$):
\begin{equation}
\frac{\oiii}{\hb} = 1.6 (\pm 0.3) \log{(L_\hb)} - 47 (\pm 9).
\label{eq:fitlmcnl}
\end{equation}
\end{itemize}

\noindent
In both cases, the slope of the SMC fit is significantly steeper than
that of LMC.
The slope of the second LMC fit (HEBs excluded,
Eq.\,\ref{eq:fitlmcnl}) is $\sim$ 3 times weaker than the slope
 of the SMC fit (Eq.\,\ref{eq:fitsmc}).

\section{Discussion}

\label{sec:discussion}
All the sample objects, except one (LMC N\,159-5), belong to the 
renowned \cite{henize56} catalog,
which presents a collection of the brightest emission nebulae in the MCs. 
Several surveys searching for planetary nebulae (PNe) have revealed that a small number of 
the catalog objects are in fact  PNe \cite[][ and references therein]{jacoby02,dopita97}.  
Therefore, we excluded the known PNe  (Table \ref{tab:resultats_misc}) from our sample. 
Consequently, we are quite confident that the sample of HEBs and LEBs does not contain 
this type of emission nebulae. As for N159-5 and, similarly, for other 
HEBs, our previous spectroscopic observations have confirmed their \hii\, region nature 
\cite[][ and references therein]{mhm99c}. Generally speaking, at the distance of MCs, 
the PNe typically subtend about 1\asec\, on the sky \citep{dopita97, stanghellini02}, 
and our inspection of the size and morphology of the objects in the available images 
confirm the smaller size and higher compactness of the PNe with respect to 
the HEBs and LEBs. 
Note that the PN sizes listed in Table \ref{tab:resultats_misc} are upper limits, since 
the binned CCD pixels degrade the profiles of point-like sources. 
The significantly smaller size and mass of ionized gas are the main reasons  
that PNe are much less luminous (compare Tables \ref{tab:resultats_LMC} and 
\ref{tab:resultats_SMC} with Table  \ref{tab:resultats_misc}). 
Thus the brightest PN in the SMC is fainter than intermediate LEBs.  \\

The distribution of the observed objects in the excitation-luminosity space
is not quite similar for the LMC and SMC samples (Fig. \ref{fig:diag}).  
The whole SMC sample can be readily fitted by a least-square line, 
as shown in Fig. \ref{fig:diag}. The LMC objects follow a 
``straightforward'' linear law up to an excitation of \oiii\,/\hb\, $\sim$\,3.5. 
Approximating the whole LMC sample by a single least-square fit requires justification. \\

The linear fits are not surprising. In fact, current photoionization models
of spherically symmetric \hii\, regions suggest that the \oiii/\hb\, ratio 
 depends on  the effective temperature of the star(s), the ionization parameter, and the gas
metallicity \citep{stasinska90,stasinska96}. The ionization parameter is defined as 
$U=Q/4\pi R^{2}nc$, where $Q$ is the total number of ionizing photons with
energy above 13.6 eV, $R$ the Str\"omgren radius, $n$ the hydrogen
density, and $c$ the speed of light.  This dimensionless parameter also depends 
on the volume filling factor via $R$ \citep{stasinska04}. 
Models by Stasi\'nska (1990)
show a linear correlation between
\oiii/\hb\, and the \hb\, luminosity for \hii\, regions with the same size, 
the same gas density, and the same number of exciting stars, but with increasing
effective temperatures. The reason for this relationship is that 
as the temperature increases, producing higher
\oiii\,/\hb\, ratios, the value of $Q$ increases as well, increasing the 
\hb\, luminosity. The models also show that lower metallicity environments favor higher
\oiii/\hb\, ratios, but the metallicity dependence can be outweighed
by the ionization parameter. \\

The observed objects in each galaxy have more or less the same physical size 
and probably the same chemical abundances. Although the gas density 
can vary from object to object,
the observed ascending behavior of the linear fits should mainly be due 
to increasing stellar temperatures.  The scatter around the least-square fits is very likely 
due to differences in ionization boundedness, density structure, local dust, 
and age. It should be underlined that this linear behavior is characteristic 
of \hii\, regions, and planetary nebulae (Table\,\ref{tab:resultats_misc}) do 
not follow this trend. In fact, although the planetary nebulae can have 
higher \oiii/\hb\, ratios, they are all underluminous with respect to \hii\, 
regions. \\

Two groups of objects show up in both galaxies. Those populating the upper right 
parts of the plots are known to be HEBs.  These are 
LMC N\,160A1, N\,11A, N\,159-5, N\,83B and SMC N\,88A, N\,81  
\citep[see][ and references therein]{mhm02a}.  
They were called so because of their small size, compactness, 
and higher \oiii/\hb\, ratio compared to common giant \hii\, regions in 
the MCs. Here, in light of new observations, it is possible 
to provide a more precise 
definition taking all the data into account. In the LMC sample,  
an HEB is an object that stands above the linear fit in 
the \oiii/\hb -{\it Log L(\hb)} 
space  (Fig. \ref{fig:diag}) and that has a luminosity of {\it Log L(\hb)} \ab\, 30.0 W or 
{\it Log L(\hb) } \ab\, 3.4 (\lsun). This means that 
an LMC HEB should at the same time have an \oiii/\hb\, ratio higher than \ab\,4.0. 
If we use the flatter fit, N\,105A-IR can also qualify as an HEB.
As regards the SMC HEBs, the ``classical'' members, N\,88A and  N\,81, 
fulfill the two criteria specified above. They have at the same time  
higher excitation compared to the LMC HEBs (see below).
 Consequently, as a working hypothesis, we call LEBs 
all the other objects that do not meet the above requirements for excitation and luminosity.
Of course one can also consider 
an intermediate group between the two extreme cases. \\

As mentioned above, Fig. \ref{fig:diag} also shows that the HEBs in 
the SMC have a higher \oiii\,/\hb\, 
ratio compared to the LMC sample. This effect is the main reason for a 
steeper slope of the fit for the SMC population, whether the whole LMC
sample is considered or not. This behavior can be explained by the fact
that, for the range of metallicities characterizing the MCs, the 
\oiii\,/\hb\, ratio becomes larger when the metallicity decreases 
because of the higher temperature of the O$^{++}$ zone \citep{stasinska02}. 
Several observational works hint at an increasing electron temperature gradient with 
decreasing O abundances as a function of the galactocentric distances 
\citep[see][ and references therein]{deharveng00}. 
There is also the possibility that low metallicity 
leads to higher effective temperatures for exciting stars, but the situation is 
not quite clear \citep[see][ for discussion]{martins05}. 
Observationally, \citet{massey04} 
find higher effective temperatures  for SMC O type stars compared to their Galactic 
counterparts. This is not supported by \citet{heap06} and \citet{bouret03}, who find no 
temperature difference between O type stars in the SMC and Galaxy. 
Theoretically, O star atmosphere models by \citet{martins_phd}, which take line-blanketing effects into 
account, show that O stars in the SMC have higher effective temperatures 
than Galactic stars. This conclusion is supported by \citet{mokiem04}. \\

The pile-up behavior of the HEBs in the LMC plot is not clearly understood at this stage. 
Why is such feature not seen for the SMC HEBs? We note that only two HEBs are present in 
the SMC plot, implying that, even if this feature existed for the SMC objects, it cannot be shown 
by the present data. Tentative explanations could be given by taking    
possible variations in the ionization parameter into account. For example, ionization-boundedness 
in the HEBs, which are  powered by hotter exciting sources or by more numerous 
sources with the same temperature. Since the HEBs have more or less the same size 
and, consequently,  comparable amounts of matter, the larger Lyman continuum photon 
fluxes cannot correspondingly create more luminous \hii\, regions. 
If this explanation is valid, we need to know the  physical conditions that  
give rise to hotter stars or to a larger number of exciting stars, when the available amount 
of gas in the initial molecular clump  is comparable to that in less luminous HEBs.   
Alternatively, the difference in the density structure/filling factor can result in 
different \oiii\,/\hb\,  ratios while comparable exciting sources are present. \\

The problem is that the exciting stars of these objects are not well-known. 
Not only are they generally out of reach of ground-based observations, which lack 
the necessary angular resolution, even our {\it HST} observations, obtained using WFPC2, have 
not been able to resolve them because they are enshrouded in gas and local dust. 
This is the case for LMC N159-5 \citep{mhm99c} and SMC N88A  \citep{mhm99b}. As for 
LMC N160A1 and LMC N83B, one star is detected in each of them \citep{mhm02a,mhm01a}, 
but an STIS-like 
spectrograph is needed to observe them in order to derive their spectral classifications. 
In the case of N11A, at least 5 stars are detected towards the blob \citep{mhm01b}, the 
main one being an O6\,V \citep{parker92}. Two stars were detected 
towards SMC N81 \citep{mhm99a}, and subsequent STIS spectroscopy has shown that they have 
spectral types O6-O8\,V \citep{mhm02b,martins04}.     \\

Another  possibility is the likely presence of dust inside the HEBs. These \hii\, regions 
are associated with relatively high amounts of local dust, since they harbor 
newborn massive stars hatching from their parental molecular clouds. 
Depending on the optical properties of dust grains, 
photon absorption by internal dust can diminish the Lyman continuum flux as if the 
exciting source(s) had a lower effective temperature, thus changing the 
ionization structure of the region. 
Similarly, the loss of ionizing photons decreases the \hb\, luminosity. The presence of 
internal dust has another implication as far as the extinction correction for the line 
fluxes is concerned. We used the usual method, which is based on the Balmer decrement 
measurement, from which the reddening is derived \citep{howarth83}. 
The basic assumption in this method is that the reddening is only due to absorption by 
the interstellar dust lying on the line of sight. However, this is certainly not true 
for the HEBs. Therefore, an appropriate extinction correction needs to resolve the 
radiation transfer equation for a medium in which gas and dust are mixed. 
Notwithstanding, this is not straightforward since 
it requires knowing the physical characteristics of the dust grains 
\citep[][ and references therein]{caplan86}. 
Therefore, the \hb\, fluxes can be underestimated. \\   

There is a noteworthy difference between the environments of HEBs and LEBs in the 
MCs.  The HEBs  occur mostly in the close vicinity of giant \hii\, regions. 
In other words, they are generally associated with molecular 
clouds and regions of active star formation. In contrast, 
LEBs are isolated objects lying far from giant \hii\, regions, 
and no conspicuous molecular clouds have been detected to be 
associated with them. However, the current LEB study suffers 
dramatically from a lack of data on various aspects of their 
environment. It would be interesting 
to find out which environmental conditions give rise to the three different 
types of \hii\, regions (giants, HEBs, and LEBs) in the MCs. 
A decreasing scale in 
the size of the parental molecular cloud may be a decisive factor. 
But then the question 
becomes: under which conditions bigger and smaller molecular clouds are formed? 
We believe that 
HEBs are younger than giant \hii\, regions and were probably formed 
from leftover parts of 
the molecular clouds that brought about the adjacent giant \hii\, regions. 
The LEBs may be older, evolved HEBs, but appropriate data 
are needed still to investigate these questions.  \\

Given the difficulty in observing the MC HEBs and LEBs, it would be 
helpful to study similar objects in our Galaxy for which higher resolution 
observations are possible. First it should be underlined that 
HEBs and LEBs are particularly important in the context of massive 
star formation 
in the MCs. To our knowledge, no studies have been undertaken yet
to find and study 
compact \hii\, regions in the vicinity of Galactic giant \hii\, regions. 
As a first step in Galactic work, it would seem helpful to 
identify this type of Galactic object to compare the physical
properties of higher and lower excitation compact \hii\, regions of 
comparable sizes. Such a study will undoubtedly constitute a 
research project on 
its own, to which enough time and effort should be devoted. 
However, we see two notable handicaps: distance uncertainties 
involved in Galactic 
\hii\, region studies and the difficulty of
measuring the total luminosity of 
Galactic 
objects extending over several arcminutes on the sky. What can be said at 
this stage is that Galactic \hii\, regions like Orion, Sh2-156, Sh2-152, Sh2-269, and 
RCW34 \citep{simpson73,mhm_sh156,mhm_sh152,mhm_sh269,mhm_rcw34}
have the same physical size and excitation as the MC LEBs. 
It remains to identify their higher excitation counterparts 
in the peripheral zones of giant \hii\, regions.

\acknowledgements{We are grateful to Dr. Gra\.zyna Stasi\'nska, of the Paris 
Observatory/CNRS, for helpful discussions and comments. 
We would also like to thank an anonymous referee for remarks and 
suggestions that helped improve the paper. }

\bibliographystyle{aa}
\bibliography{biblioleb}

\end{document}